\documentstyle[seceq,epsf]{ptptex}

\title{Kaonic Nuclear Systems $\bar{K} N$ and $\bar{K} NN$ as Decaying States}

\author{Yoshinori {\sc Akaishi}$^{*,***}$, Khin Swe {\sc Myint}$^{**}$ 
        and Toshimitsu {\sc Yamazaki}$^{*,****}$}

\inst{$^*$RIKEN Nishina Center, Wako, Saitama 351-0198 \\
$^{**}$Mandalay University, Mandalay, Union of Myanmar \\
$^{***}$College of Science and Technology, Nihon University, Funabashi, Chiba 274-8501 \\
$^{****}$Department of Physics, University of Tokyo, Tokyo 113-0033}

\recdate{ 
\today
}

\abst{The formation spectra of model $\bar KN$ and $\bar KNN$ systems formed by $(K^-, n)$ reactions are investigated in order to obtain a theoretical basis for a proper interpretation of experimental data concerning kaonic nuclear quasi-bound states. It has been clarified that the experimentally observable kaonic nuclear state $K^-pp$ should be regarded as the decaying state introduced by Kapur-Peierls, which is different from the pole state solution of the Faddeev equation.}

\begin{document}

\maketitle     

~~~~~~~~~~~[ To be published in Proc. Japan Acad. B ]

\section{Introduction}
	
In recent years we have predicted deeply bound kaonic states, and studied their structure and formation\cite{Akaishi02,Yamazaki02,Dote04a,Dote04b,Yamazaki04,Akaishi05,Kienle06,Yamazaki07,Yamazaki07a,Yamazaki07b} based on the $\bar{K} N$ interaction, which was derived by a coupled-channel calculation so as to account for the empirically known low-energy $\bar{K} N$ quantities. The predicted states have shown astonishing properties, such as deep binding and high nuclear densities. Among them the most basic is the $K^- pp$ system, which was predicted to be a quasi-bound state of 48 MeV binding and 60 MeV width\cite{Yamazaki02} by a variational calculation using a complex $\bar KN$ interaction. This system has been fully studied,\cite{Yamazaki07a,Yamazaki07b} revealing that a super-strong nuclear force is caused by a $\bar{K}$, which migrates in a dynamically formed molecular-type dense structure. Lately, coupled-channel Faddeev calculations have been done for the same $K^-pp$ system,\cite{Shevchenko07a,Ikeda07,Shevchenko07b} but the binding energy ranges over $\sim 50 - 80$ MeV with a much larger width of $\sim 100$ MeV. Note that the widths calculated by all of these different authors are the partial widths for the pionic decay modes of $\rightarrow \pi \Sigma p~(>85\%)$ and $\rightarrow \pi \Lambda p~(<15\%)$, the former of which is closed when the $K^-pp$ binding energy exceeds about 100 MeV. Shevchenko-Gal-Mare$\check{\rm s}$-R$\acute{\rm e}$vai\cite{Shevchenko07b}  criticized the use of an energy-independent complex $\bar KN$ interaction by Yamazaki-Akaishi (Y-A), to which they attributed the origin of the discrepancy of the predicted widths. In the present paper we consider this problem, and clarify that the pole solution of the Faddeev equation does not correspond to an experimentally observable physical quantity, whereas the treatment of Y-A effectively takes into account the decaying process realistically. 

The paper is organized as follows. First, we consider a model $\bar K N$ quasi-bound state by changing the strength of the $\bar K N$ interaction, and clarify the difference between the pole state and the decaying state in a formation reaction by introducing the concept of Kapur-Peierls.\cite{Kapur38} Then, we proceed to the $K^-pp$ system, where we show that the smaller width in our treatment arises from an effective consideration of the realistic decaying process in contrast to the solution of the Faddeev equation.  As experiments dedicated to the issue on the existence of such kaonic nuclei are planned at DA$\Phi$NE, GSI and J-PARC, relevant theoretical framework should be carefully checked and developed. In this context we propose the concept of an "intrinsic decaying state" to interpret experimental data of deeply bound $\bar K$ states.

\section{Formation of a model $\bar K N$ quasi-bound state}

\subsection{Solvable model setting}

We start from the assumption that the $\Lambda (1405)$ resonance is an $I=0$ quasi-bound state of $\bar KN$, which is embedded in continuum of $\Sigma \pi$ as a kind of Feshbach resonances\cite{Feshbach58}. In general, quasi-stable bound states of an exotic hadronic particle, such as the present kaonic bound states, the deeply bound pionic states, and the metastable antiprotonic helium states, are characterized as special kinds of Feshbach resonances, where new hadronic particles are born at high excitation and reveal themselves as bound states near their emission thresholds.\cite{Yamazaki:00}  They are all embedded in continuum, but persist to be discrete states. Among them, the $\bar KN$ resonance state is the simplest system to study its physics deeply.

We consider two channels of $\bar KN~(K^-p)$ and $\pi\Sigma~(\pi^-\Sigma^+)$ for simplicity. We  employ a set of separable potentials with a Yukawa-type form factor\cite{Yamaguchi54}, 
\begin{eqnarray}
\langle \vec k' \mid v_{ij} \mid \vec k \rangle = g(\vec k') ~U_{ij} g(\vec k), ~~~g(\vec k) = \frac {\Lambda^2} {\Lambda^2 + \vec k^2}, \\
U_{ij} = \frac{1}{\pi^2} \frac{\hbar^2}{2 \sqrt{\mu_i \mu_j}}
\frac{1}{\Lambda} s_{ij}, \hspace{2cm}
\label{YYint}
\end{eqnarray} 
where $i(j)$ stands for the $\bar KN$ channel, 1, or the $\pi \Sigma$ channel, 2, $\mu_i(\mu_j)$ is the reduced mass of the channel $i(j)$ and $s_{ij}$ are non-dimensional strength parameters. The binding energy, $B_{\bar K} = 27$ MeV, and the width, $\Gamma = 40$ MeV, of $\Lambda(1405)$ are reproduced with the values of
\begin{equation}
s_{11}^{(0)}=-1.288,~s_{12}=0.2783,~s_{22}=-0.660,
\label{strength}
\end{equation}
where $U_{22}/U_{11}^{(0)} = 4/3$, like in a "chiral" model, and $\Lambda = 770$ MeV$/\hbar c = 3.90$ fm$^{-1}$ are adopted.

\begin{figure}[hbtp]
\begin{center}
\hspace{0.5cm}
\vspace{-0.5cm}
\epsfxsize=5.5cm
\epsfbox{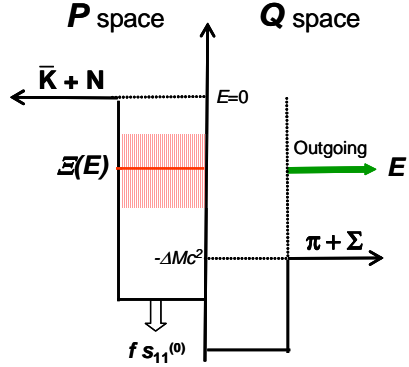}
\vspace{0.5cm}
\end{center}
\caption{Schematic picture of the present model, which has a quasi-bound state of $\bar KN$ with complex energy $\Xi (E)$, decaying to $\pi \Sigma$ with energy $E$, which is generally complex. The pole state of the coupled system satisfies $\Xi (E) = E$.}
\label{model}
\end{figure}

Our theoretical interest is how the excitation spectrum of the quasi-bound state, $\Lambda^*(\bar KN)$, behaves when the bound state comes closer to the $\pi\Sigma$ lowest decay threshold. In order to investigate the spectrum shape we increase the attractive strength of the $\bar KN$ channel interaction as 
\begin{equation}
s_{11}^{(0)} \rightarrow s_{11} = f \cdot s_{11}^{(0)}.
\end{equation}
The model is depicted in Fig.~\ref{model}.
%

The coupled-channel Schr\"odinger equation is written with Feshbach's projection operators, $P$ and $Q$ $(\equiv 1 - P)$, to the subspaces $P$ and $Q$, as  
\begin{eqnarray}
PHP~P\Psi + PVQ~Q\Psi = E~P\Psi, \\
QHQ~Q\Psi + QVP~P\Psi = E~Q\Psi,
\label{cceq}
\end{eqnarray}
where $H = T + V$ is the coupled-channel Hamiltonian.\cite{Feshbach58} In the $P$ space, Green's function holds the following relation:\cite{Morimatsu85}
\begin{equation}
P \frac {1}{E-H+i\epsilon} P = P \frac {1}{E-H^{\rm{opt}}+i\epsilon} P,
\label{prop}
\end{equation}
where the "optical" Hamiltonian is defined by
\begin{equation}
H^{\rm {opt}} = PHP + PVQ \frac {1}{E-QHQ+i\epsilon} QVP.
\label{Hopt}
\end{equation}
The complex potential thus derived corresponds to the generalized optical potential in a standard nuclear reaction theory, and thus we refer to it as {\it optical potential} hereafter. It should be emphasized that this procedure (and the thus-derived complex potential) in the present case of coupled $K^-p$ and $\pi\Sigma$ channels leads to exact outcomes in $P$ space, while such a single-channel complex potential is sometimes misunderstood as being a crude approximation.\cite{Shevchenko07b} The solution using the above complex potential is totally equivalent to the solution of a direct coupled-channel treatment. 

In the case of the present model the optical potential in the $\bar KN$ channel as a function of the complex energy $E$ measured from the $K^- + p$ threshold is given by 
\begin{equation}
v_{1}^{\rm {opt}}(E) = v_{11} + v_{12} \frac {1}{E-h_{22}+i\epsilon} v_{21}
\label{vopt}
\end{equation}
with $h_{22} = t_2^{\rm{kin}} + v_{22} - \Delta Mc^2$, where $\Delta M = m_{K^-} + M_p - m_{\pi^-} - M_{\Sigma^+} =103$ MeV$/c^2$ is the threshold mass difference. The corresponding optical strength is analytically derived as shown in Appendix to be
\begin{equation}
s_{1}^{\rm {opt}}(E) = s_{11}-s_{12} \frac {\Lambda^2}{(\Lambda-i\kappa_2)^2+s_{22}\Lambda^2} s_{21}, ~~ \frac {\hbar^2}{2 \mu_2} \kappa_2^2 = E+\Delta Mc^2,
\label{sopt}
\end{equation} 
where $\kappa_2$ is a complex momentum in the $\pi \Sigma$ channel. The {\it et al}gy, $E_{\rm{pol}}$, of the quasi-bound {\it pole} state is obtained by satisfying $E_{\rm{pol}} = \Xi (E_{\rm{pol}})$,
where 
\begin{equation}
\Xi (z) \equiv - \frac {\hbar^2}{2 \mu_1} \Lambda^2 (\sqrt{- s_1^{\rm {opt}}(z)} - 1)^2, 
\label{E0}
\end{equation} 
which is the eigen-value of $h_1^{\rm{opt}}(z) = t_1^{\rm{kin}} + v_1^{\rm{opt}}(z)$ on a proper Riemann's sheet.

\begin{figure}[hbtp]
\begin{center}
\hspace{0.5cm}
\vspace{0cm}
\epsfxsize=14cm
\epsfbox{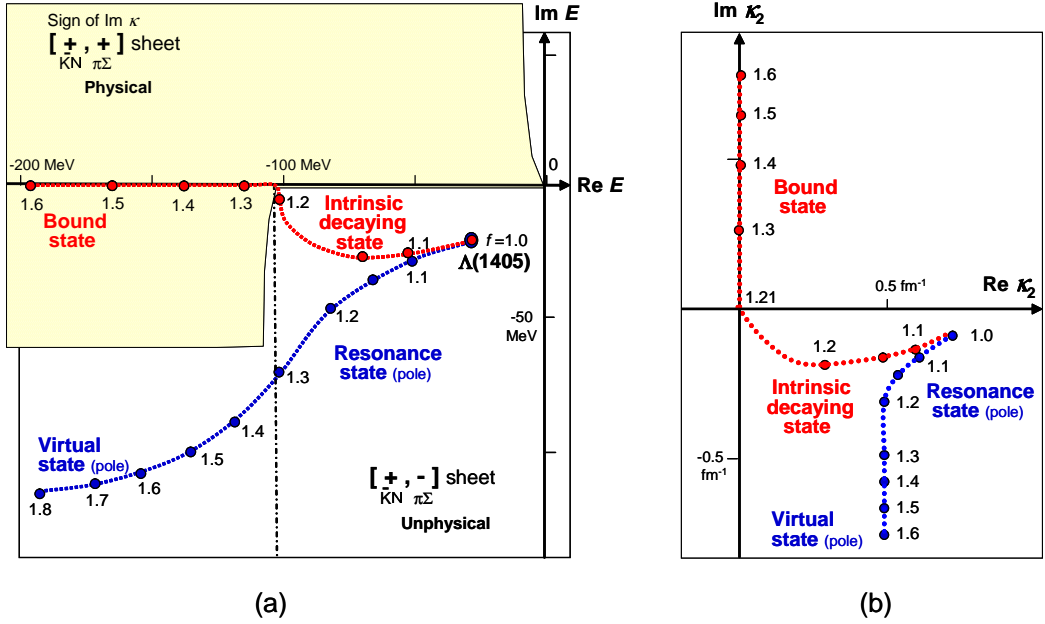}
\vspace{1.0cm}
\end{center}
\caption{Trajectory of the pole on the $E$ plane, (a), and on the $\kappa_2$ plane, (b), when the strength of the $\bar KN$ interaction is increased as $s_{11}^{(0)} \rightarrow f \cdot s_{11}^{(0)}$ with $f =1.0 - 1.8$. The pole state close to the $\pi\Sigma$ threshold deviates from the experimentally expected behavior, in contrast to the intrinsic decaying state defined by Eq.(\ref{intrinsic}). The symbol $+$ ($-$) denotes the sign of Im $\kappa_1$ or Im $\kappa_2$ in Riemann's physical (unphysical) sheet.}
\label{Poles}
\end{figure}

The pole of the present dynamical system moves with increasing $f$ as shown in Fig.~\ref{Poles}. The pole state, as it becomes deeper, deviates from the experimentally expected behavior. Namely, the width becomes broader and broader toward the $\pi\Sigma$ decay threshold, and this tendency persists even beyond this kinematical limit. At the threshold the pole state goes to a virtual state, not to a bound state of $\pi\Sigma$. Thus, the pole state has an unreasonable behavior, when it is broad and close to the threshold. This situation arises from ignorance about the on-shellness of the decaying particles. Then, how can we describe experimentally observable states? In the next subsection we will introduce "intrinsic decaying state", imposing the on-shell condition to the outgoing particles. Its behavior is also shown in Fig.~\ref{Poles}. The state becomes narrower and narrower toward the $\pi\Sigma$ threshold, exhibits a sharp cusp at around $f=1.2$ just before the threshold, and turns into a stable bound state by changing Riemann's sheet from $[+,-]$ to a $[+,+]$ physical one at the $\pi\Sigma$ threshold, where the first and second signs are those of Im $\kappa$ for $\bar KN$ and $\pi\Sigma$, respectively.

\subsection{Spectra of the pole state and the decaying state}

\begin{figure}[hbtp]
\begin{center}
\hspace{0.5cm}
\vspace{0.0cm}
\epsfxsize=9.7cm
\epsfbox{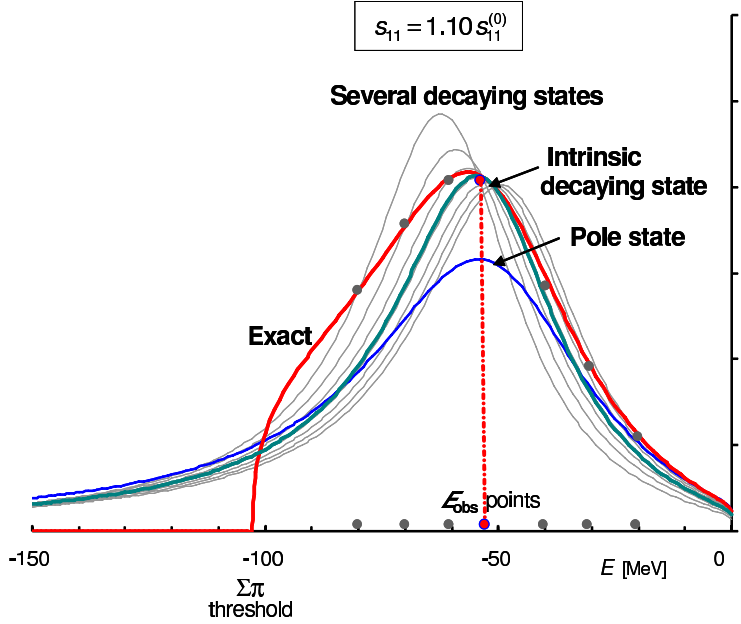}
\vspace{0cm}
\end{center}
\caption{Spectrum shapes of decaying states in the case of $1.10~s_{11}^{(0)}$. The intrinsic decaying state has 52 MeV binding, as denoted by the dot-dashed vertical line, and 50 MeV width. The exact spectrum is obtained by connecting the spectrum values of the decaying states at respective $E_{\rm{obs}}$ points. The pole state is also shown for a comparison.}
\label{decaying}
\end{figure}

\begin{figure}[hbtp]
\begin{center}
\hspace{0.5cm}
\vspace{0cm}
\epsfxsize=9.7cm
\epsfbox{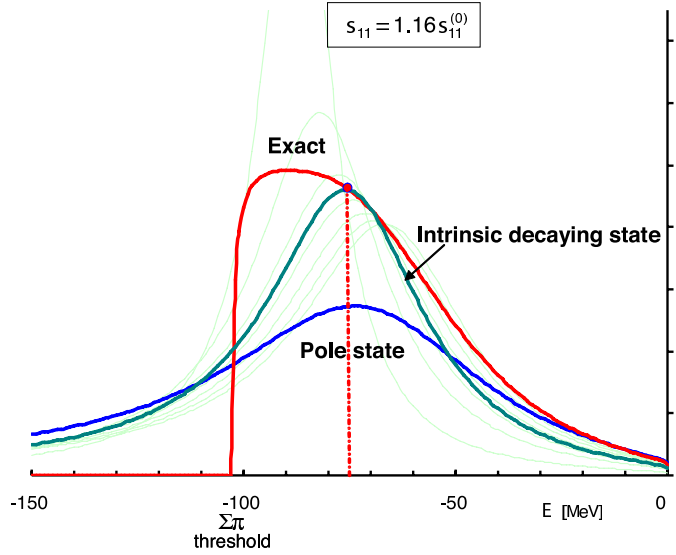}
\vspace{0cm}
\end{center}
\caption{Spectrum shapes of the intrinsic decaying and the pole states together with an exact one in the case of $1.16~s_{11}^{(0)}$. The intrinsic decaying state provides a better description of the spectrum than does the pole state.}
\label{spec}
\end{figure}

\begin{figure}[hbtp]
\begin{center}
\hspace{0.5cm}
\vspace{-0.5cm}
\epsfxsize=9.7cm
\epsfbox{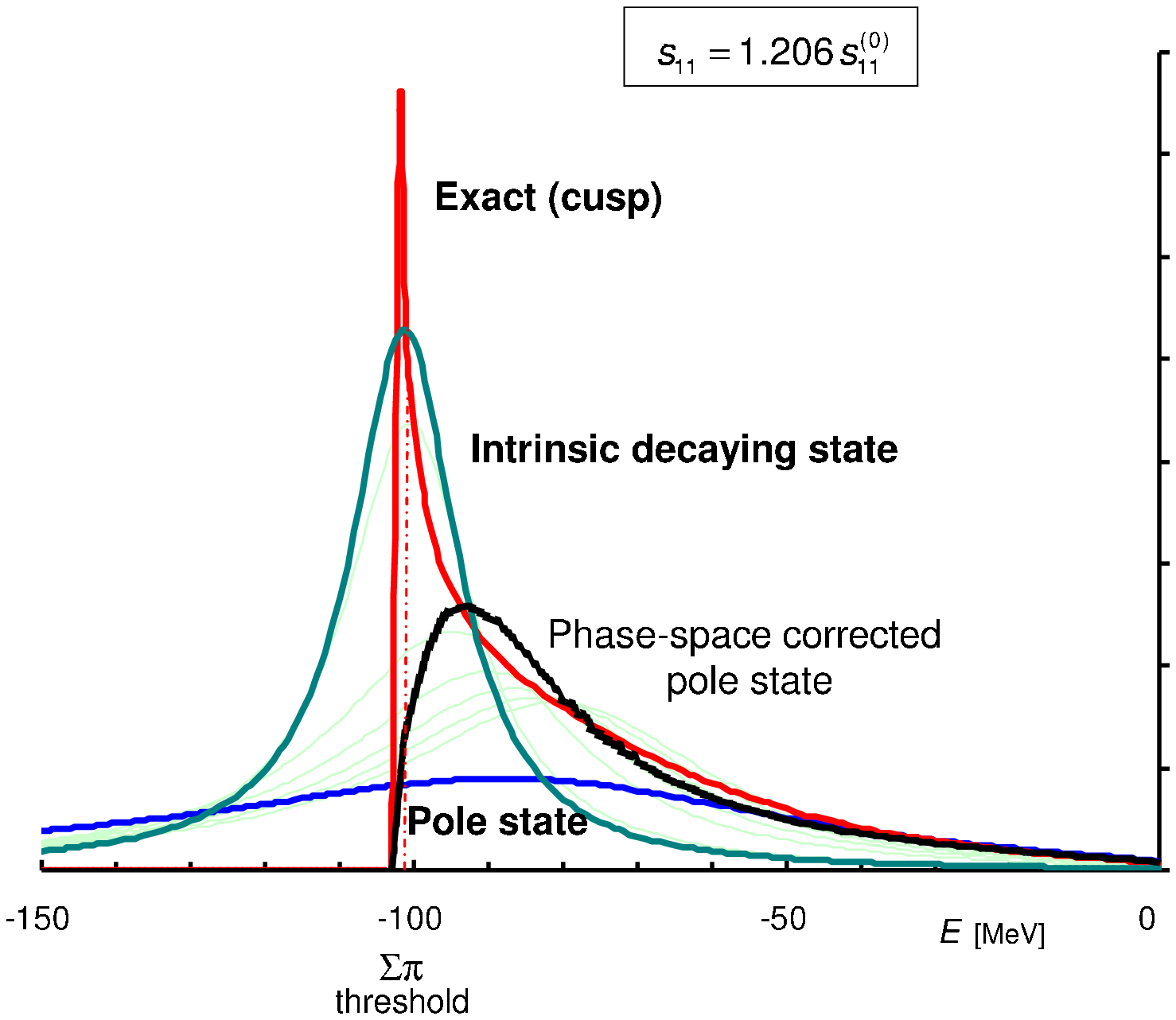}
\vspace{0.5cm}
\end{center}
\caption{Spectrum shapes of the intrinsic decaying and the pole states together with the exact one in the case of $1.206~s_{11}^{(0)}$. The intrinsic decaying state provides a far better description of the spectrum than the pole state and its corrected one do, in the cusp case.}
\label{cusp}
\end{figure}

We now consider how to form such a model quasi-bound state by a representative reaction 
\begin{equation}
K^- + d \rightarrow \Lambda^*(\bar KN) + n \rightarrow \pi^- + \Sigma^+ + n.
\label{Kdreac}
\end{equation}
The missing-mass spectrum from this reaction is calculated by using Green's function,\cite{Morimatsu85} as follows:
\begin{eqnarray}
\frac{d^3\sigma}{d\vec k_n} = (2\pi)^4 \frac{E_K^{\rm{in}}}{\hbar^2c^2k_K^{\rm{in}}} \vert \langle \vec k_n \mid \Phi_d \rangle \vert^2  ~(-\frac{1}{\pi}) ~{\rm{Im}} ~\big\lbrack \int d\vec r' \int d\vec r \hspace{0.5cm} \nonumber \\
\times ~\langle \vec k_{\rm{rel}}^{\rm{in}} \mid t^\dagger \mid \vec r' \rangle \langle \vec r' \mid \frac{1}{E-h_{1}^{\rm{opt}}(E)+i\epsilon} \mid \vec r \rangle \langle \vec r \mid t \mid \vec k_{\rm{rel}}^{\rm{in}} \rangle ~\big\rbrack ,
\label{cross}
\end{eqnarray} 
where $E_K^{\rm{in}}, k_K^{\rm{in}}$ are the incident energy and momentum of $K^-$, $\Phi_d$ is a deuteron wave function, and $\langle \mid t \mid \rangle$ is a transition matrix between $K^-$ and $p$ in $d$ from the initial state with a relative momentum $\vec k_{\rm{rel}}^{\rm{in}}$ to final states in quasi-bound region which we are interested in. The quantity $E$ is the missing mass multiplied by $c^2$ of the $\bar KN$ system, which is a {\it real}-value variable depending on the kinematical condition of the experiment. As stressed before, the use of the optical potential, Eq.(\ref{sopt}), in the calculation of Eq.(\ref{cross}) gives exactly the same spectrum as that obtained by solving the original coupled-channel Schr\"odinger equation, thanks to the relation of Eq.(\ref{prop}). This exact spectrum is shown in Figs.~\ref{decaying}, \ref{spec} and \ref{cusp}. The spectrum, which depends on the imaginary part of $h_{1}^{\rm{opt}}(E)$, vanishes below the $\pi\Sigma$ threshold, as expected, since the optical potential of Eq.(\ref{sopt}) changes from complex to real one due to the purely imaginary $\kappa_2$. The procedure of Eq.(\ref{cross}) using $h_{1}^{\rm{opt}}$ is essentially a calculation of the "decaying state" of the system.

The {\it decaying state} was introduced by Kapur-Peierls\cite{Kapur38} as an eigen-state formed in the "internal region" properly limited, from which outgoing {\it on-shell} particles emerge in the asymptotic region of open channels. The complex energy (position and width) of the decaying state is given by using Eq.(\ref{E0}) as 
\begin{equation}
E_{\rm{dec}} = \Xi (E_{\rm{obs}})
\label{Edec}
\end{equation}
for a real energy $E_{\rm{obs}}$ of each measurement point of the experiment.
%
A Breit-Wigner type spectrum, 
\begin{equation}
S^{\rm BW} (E~;E_{\rm dec}) = \frac{1}{\pi} \frac{-~{\rm Im}~E_{\rm dec}}{(E-{\rm Re}~E_{\rm dec})^2 + ({\rm Im}~E_{\rm dec})^2},
\end{equation}
 can be drawn for each value of $E_{\rm obs}$ from Eq.(\ref{cross}) as a one-level formula. Figure \ref{decaying} (thin curves) shows such spectra for several values of $E_{\rm{obs}}$. The real part of $E_{\rm dec}$ generally deviates from $E_{\rm obs}$, and each Breit-Wigner curve has a crossing point at $E=E_{\rm obs}$, as indicated by a dot. The locus of such dots shows a smooth curve, expressed by
 \begin{equation}
S(E_{\rm obs}) = S^{\rm BW} (E_{\rm obs}; E_{\rm dec}),
\end{equation}
which is found to be equivalent to the exact spectrum calculated by 
 replacing $h_{1}^{\rm{opt}}(E)$ by $h_{1}^{\rm{opt}}(E_{\rm{dec}})$ in the Green function of Eq.(\ref{cross}). The decaying state is not a unique state, but an ensemble of states; it depends on the variable $E$, which changes under energy-momentum conservation in the measurement. To overcome such complexity, we introduce the "intrinsic decaying state" designed to be a representative eigen-state of a school of decaying states.

%
%
The {\it intrinsic decaying state} is defined as an eigen-state with the complex eigen-energy, $E_{\rm dec}^{\rm int}=\Xi$, that satisfies the equation of
\begin{equation}
z - {\rm Re} ~\Xi (z) = 0, 
\label{intrinsic}
\end{equation} 
whereas the energy of the pole state, $E_{\rm pol}=\Xi$, is a solution of
\begin{equation}
z - \Xi (z) = 0.
\label{pole}
\end{equation} 
These equations impose consistency between the boundary condition ($z$) and the eigen-value ($\Xi $). In the case of Eq.(\ref{intrinsic}) the parameter $z$ becomes a real number and assures the on-shellness of decay particles incorporated into $h_1^{\rm{opt}}(z)$. Since $h_1^{\rm{opt}}(z)$ itself is of complex, the eigen-value $\Xi(z)$ is a complex number, and thus a complex $E_{\rm dec}^{\rm int}$ is obtained as a consistent solution. The complex eigen-values, $E_{\rm dec}^{\rm int}$ and $E_{\rm pol}$, of the two different states are seen for various values of $f$ in Fig.~\ref{Poles}. The spectrum of the intrinsic decaying state is obtained with BW function as $S^{\rm BW} (E~;E_{\rm dec}^{\rm int})$, whereas that of the pole state is given as $S^{\rm BW} (E~;E_{\rm pol})$. Those spectra including a continuum are obtained by replacing $h_{1}^{\rm{opt}}(E)$ by $h_{1}^{\rm{opt}}(E_{\rm dec}^{\rm int})$ and $h_{1}^{\rm{opt}}(E_{\rm pol})$, respectively, in the Green function of Eq.(\ref{cross}).

The case of $f=1.16$ is shown in Fig.~\ref{spec}. The pole-state spectrum has a long tail below the lowest $\pi\Sigma$ decay threshold, which does not satisfy the kinematical condition, and thus cannot be observed by any experiment. Now, one should notice that an experimental observation corresponds {\it not to the "pole state", but to the "decaying state"}, since the detectable decay particles, $\pi$ and $\Sigma$, appear in the asymptotic region as on-shell objects. The intrinsic decaying state gives a much better description of the spectrum than does the pole state in the case of a broad deeply bound state. The energy and the width of $\bar K$ are Re $E_{\rm pol} = -70$ MeV and $\Gamma_{\rm pol} = 74$ MeV for the pole state, and Re $E_{\rm dec}^{\rm int} = -75$ MeV and $\Gamma_{\rm dec}^{\rm int} = 45$ MeV for the intrinsic decaying state. The width of the intrinsic decaying state is considerably smaller than that of the pole state.

Figure \ref{cusp} shows the case with $f=1.206$, where the exact spectrum reveals a sharp cusp. The pole-state spectrum persists to be broad, while the intrinsic  decaying state again describes the spectrum far better than does the pole state; Re $E_{\rm pol} = -83$ MeV and $\Gamma_{\rm pol} = 97$ MeV for the pole state, and Re $E_{\rm dec}^{\rm int} = -101$ MeV and $\Gamma_{\rm dec}^{\rm int} = 18$ MeV for the intrinsic decaying state. A modification of such a broad pole-state spectrum has been proposed to multiply the imaginary part of the optical potential by a phase-space weight of the decay channel.\cite{Mares06}. A modified spectrum is also compared in Fig.~\ref{cusp} as "phase-space corrected pole state", but it does not reproduce the cusp structure.

\subsection{Properties of the decaying state}

\begin{figure}[hbtp]
\begin{center}
\hspace{0.5cm}
\vspace{-0.7cm}
\epsfxsize=12cm
\epsfbox{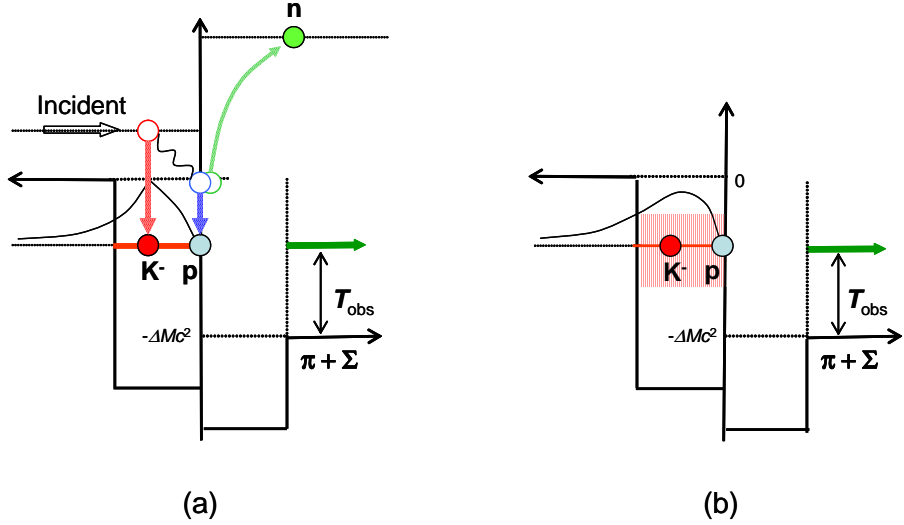}
\vspace{0.5cm}
\end{center}
\caption{Schematic picture for the decaying state formed in the reaction $K^- + d \rightarrow K^- p + n$. (a) A stationary state with  real energy is formed over all of the channels related to the reaction process. (b) A transient state with complex energy $\Xi (T_{\rm{obs}}-\Delta M c^2)$, which is the decaying state, is formed in the $\bar KN$ channel under the boundary condition of on-shell decay to the $\pi\Sigma$ open channel.}
\label{scheme}
\end{figure}

Figure \ref{scheme} gives a schematic picture for the process of the reaction Eq.(\ref{Kdreac}) and the decaying state. In the reaction process of the left panel (a) all of the energies of the incident $K^- d$, the $K^-p$ and the $\pi\Sigma$ channels have real values. In the $K^-p$ channel the real energy is assured by a feeding from the incident channel, as discussed below. First we consider a single feeding point case. Then, the $K^-p$ wave function obeys the following equation with a source term:
\begin{equation}
-\frac {\hbar^2}{2\mu_1} \frac {d^2}{dr^2} u_1(r) + [v^{\rm{opt}} u_1](r) - E u_1(r) = -V^{\rm{feed}}(r_0)~\delta (r-r_0),
\label{source} 
\end{equation} 
where $E=T_{\rm{obs}}-\Delta Mc^2$ is any real energy kinematically allowed. To satisfy the boundary conditions of $u_1(0)=0$ and $u_1(\infty)=0$ the wave function must have a kink at $r=r_0$, the strength of which is defined as
\begin{equation}
a(E)=\big[ \frac{du_1}{dr}\big\vert_{r_0 -0}-\frac{du_1}{dr}\big\vert_{r_0+0} \big] \big/ u_1(r_0). 
\label{kink}
\end{equation} 
Then, the following relation is obtained: 
\begin{equation}
\vert u_1(r_0) \vert^2 = \big\vert \frac{2\mu_1}{\hbar^2} \frac{V^{\rm{feed}}(r_0)}{a(E)} \big\vert^2. 
\label{population}
\end{equation} 
This means that the weaker is the kink the stronger is the population of the state, since the excitation strength by the reaction is proportional to $\vert u_1(r_0) \vert^2$. The experimentally observable spectra shown as "exact" in Figs.~\ref{decaying}~-~\ref{cusp} are distributions of this kind of "population strength". The above argument is easily extended to a general case by integrating over $r_0$, using $\int V^{\rm{feed}}(r_0) \delta (r-r_0) dr_0 = V^{\rm{feed}}(r)$.

When the feeding is cut off, a transient state is formed in the $K^-p$ channel with a complex energy due to the decay to the $\pi\Sigma$ open channel, which is the decaying state sketched in the right panel (b) of Fig.~\ref{scheme}. Since the total energy of decaying particles, $\pi$ and $\Sigma$, has been kinematically determined through an experimental process, the energy of this transient state is not dispersive, but is uniquely given. Then, the imaginary part of the energy denotes the decay width, $\Gamma = \hbar / \tau_{\rm{life}}$, giving information about the lifetime, $\tau_{\rm{life}}$, of the transient state. As is understood from Fig.~\ref{decaying} the experimentally obtained width (missing-energy spread) is not the decay width, but the distribution width of the population strength. Strictly speaking, both widths coincide only when a common decaying state is formed for any $T_{\rm{obs}}$ in the peak region. In an actual case, in order to know the lifetime of a quasi-bound/resonance state one must extract a Breit-Wigner type spectrum of the intrinsic decaying state through an analysis of the experimental data. Figures \ref{decaying}, \ref{spec} and \ref{cusp} demonstrate three examples of the difference between the decay width ("intrinsic decaying state") and the distribution width ("exact").
The wave function of the decaying state is explicitly obtained to be
\begin{eqnarray}
u_1(r)=N\sqrt{\mu_1} \frac {\Lambda-i\kappa_1}{\Lambda+i\kappa_1}\{e^{-\Lambda r}-e^{i\kappa_1 r}\}, ~~~~E_{\rm{dec}}=\frac {\hbar^2}{2 \mu_1} \kappa_1^2, 
\label{decayfn1} \\
u_2(r)=N\sqrt{\mu_2} \frac {-s_{21}\Lambda^2}{(\Lambda-ik_2)^2+s_{22}\Lambda^2} \frac {\Lambda-ik_2}{\Lambda+ik_2}\{e^{-\Lambda r}-e^{ik_2r}\},
\label{decayfn2}
\end{eqnarray} 
where the second one should be regarded as the boundary condition of the outgoing state with a real kinetic energy, $T_{\rm{obs}}=(\hbar k_2)^2 / (2\mu_2)$.

%

Needless to say that the eigen-state of the isolated coupled-channel system is the pole state that is formed all over the $\bar KN$ and $\pi\Sigma$ channels. However, the large difference between the exact spectrum and the pole state spectrum means that the pole state is strongly disturbed and rearranged, when the system is connected to the incident channel of the production reaction under the on-shell decay condition to the open channel. Thus, it is not effective to consider the pole state to be an entity that corresponds to the experimental peak, especially in a broad and near-threshold resonance case. We should specify our question: what structure is formed in the $\bar KN$ channel when a broad peak is experimentally observed in the $\bar KN$ missing-mass spectrum? Then, we come to a legitimate answer that the structure is the "intrinsic decaying state" as a representative of decaying states.

\section{Decaying state and pole state of $K^-pp$}

\subsection{Energy dependence of the complex $\bar KN$ interaction}

\begin{figure}[hbtp]
\begin{center}
\hspace{0.5cm}
\vspace{-0.5cm}
\epsfxsize=13cm
\epsfbox{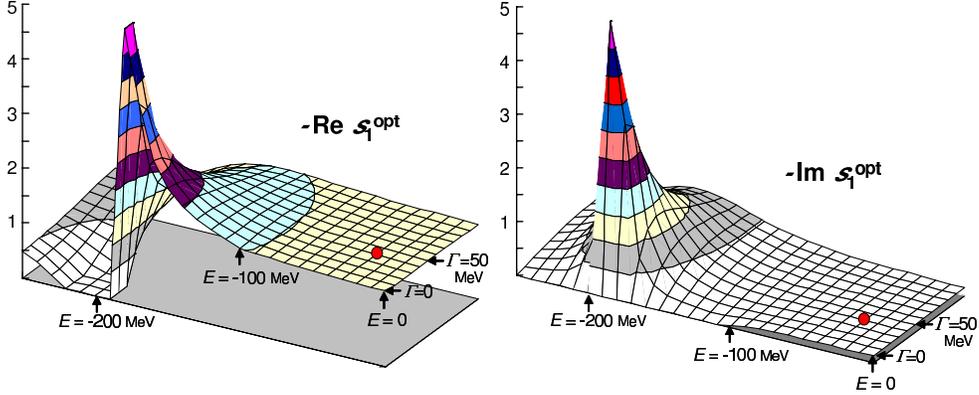}
\vspace{0.5cm}
\end{center}
\caption{Behavior of $s_1^{\rm{opt}}(z)$ as a function of $z=E-i\Gamma/2$ on Riemann's $[+,-]$ sheet. The circle denotes the position of $\Lambda(1405)$.}
\label{vplot}
\end{figure}

Now let us extend our viewpoint and discussion to the most basic kaonic nucleus, $K^-pp$. This system was predicted by using an energy-independent complex $\bar KN$ potential,\cite{Yamazaki02,Yamazaki07b} which was determined phenomenologically from the $\Lambda(1405)$ state and the $\bar KN$ scattering length.\cite{Akaishi02}  In order to understand its theoretical background we investigate the energy dependence of the optical strength of Eq.(\ref{sopt}) for the case of the strength Eq.(\ref{strength}). Figure \ref{vplot} shows an overview of the complex strength, $s_1^{\rm{opt}}$, of the single-channel $\bar KN$ interaction as a function of $z=E-i\Gamma/2$ on Riemann's $[+,-]$ sheet (see Fig.~\ref{Poles}). A singularity appears at
\begin{equation}
z = - \frac {\hbar^2}{2 \mu_2} \Lambda^2 (\sqrt{- s_{22}} - 1)^2 - \Delta Mc^2,
\end{equation}
giving the gross structure of the energy dependence. Sometimes the singularity brings a serious energy dependence in the case of a small $\Lambda$ value.

\begin{figure}[hbtp]
\begin{center}
\hspace{0.5cm}
\vspace{-0.5cm}
\epsfxsize=10cm
\epsfbox{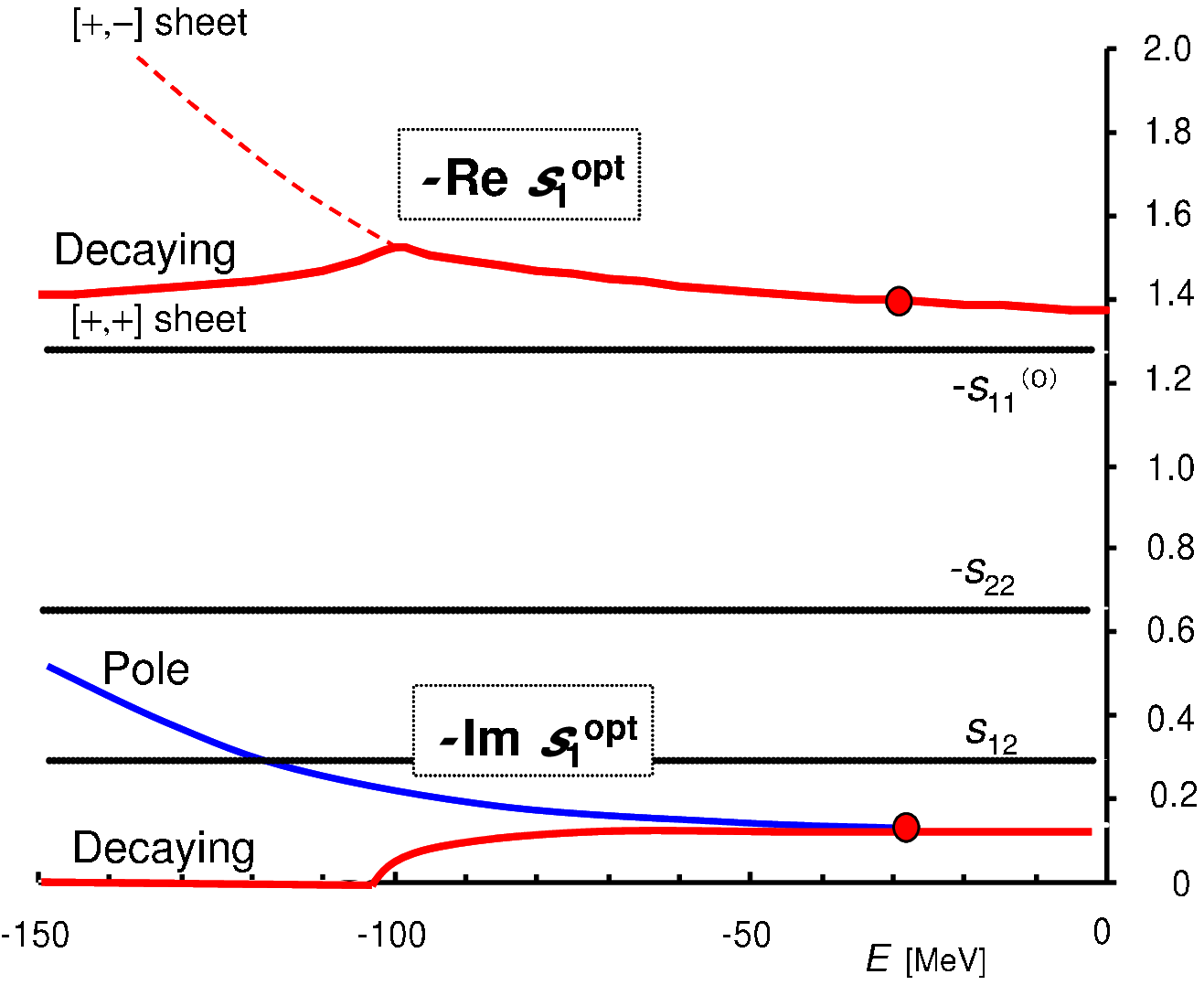}
\vspace{0.5cm}
\end{center}
\caption{Energy dependence of the optical $\bar KN$ interaction for the decaying state along the $\Gamma=0$ line. The decaying state changes Riemann's sheet from $[+,-]$ to $[+,+]$ at the $\pi\Sigma$ threshold, yielding a moderate energy dependence of Re $s_1^{\rm{opt}}$ and vanishing Im $s_1^{\rm{opt}}$ below the threshold. The broken curve shows Re $s_1^{\rm{opt}}$ on the persistent $[+,-]$ sheet. For a comparison, the imaginary part of the $\bar KN$ interaction for the pole state along the $\Gamma=100$ MeV line of Fig.\ref{vplot} is also shown.}
\label{edep}
\end{figure}

Figure \ref{edep} shows the energy dependence of the $\bar KN$ interaction to be used in a calculation of the intrinsic decaying state of $K^-pp$, which is the dependence along the $\Gamma=0$ line in Fig.~\ref{vplot}, since $\pi$ and $\Sigma$ come out as on-shell decay particles with real energies. The imaginary strength gradually becomes weaker as $E$ becomes lower and vanishes below the $\pi\Sigma$ threshold, in accordance with the physical intuition. 
%
%
%
%
On the other hand, the energy dependence of the $\bar KN$ interaction for the pole state is seen along the $\Gamma \neq 0$ path. As an example, the imaginary part along the $\Gamma=100$ MeV line is demonstrated in Fig.~\ref{edep}, which continues to grow irrespective of the $\pi\Sigma$ threshold as $E$ lowers. This is the origin of the large width of the pole state of $K^-pp$.

The real part of the $\bar KN$ interaction, Re $s_1^{\rm{opt}}$, for the intrinsic decaying state of $K^-pp$ is plotted in Fig.~\ref{edep}. Its strength has a downward kink at the $\pi\Sigma$ threshold which makes the overall energy dependence moderate. It is to be noted that the kink comes from the fact that the intrinsic decaying state changes the Riemann sheet at the $\pi\Sigma$ threshold from $[+,-]$ to $[+,+]$, as depicted in Fig.~\ref{Poles}. On the other hand, the corresponding strength for the pole state, as seen in Fig.~\ref{vplot}, increases monotonically with an upward kink at the threshold (the broken curve in Fig.~\ref{edep}), which arises from the ignorance of the change of the Riemann sheet at the threshold. Thus, {\it the use of an energy-independent complex $\bar KN$ interaction by Y-A\cite{Yamazaki02,Yamazaki07b} is not only justified, but  also found to be effectively a good approximation to obtain the intrinsic decaying state of $K^-pp$}.

\subsection{Width of the deeply bound $K^-pp$}

We want to know how much the width differs between the pole state and the intrinsic decaying state of $K^-pp$, when the states become sufficiently deep. First we set the pole state of $K^-pp$ to have Shevchenko {\it et al}.'s values of the Faddeev solution, $B(K^-pp) = 75$ MeV and $\Gamma = 100$ MeV, and determine the strengths, $s_{11}, s_{12}$ and $s_{22}$, by reproducing them. Recently, the structure of $K^-pp$ was investigated in detail,\cite{Yamazaki07a,Yamazaki07b} and it is revealed that the $K^-$ migrates between the two protons and interacts with one of them almost exclusively by  virtue of the strong $I$=0 $\bar KN$ interaction. By fully taking into account this fact, we construct the $\bar KN$ optical interactions from Eq.(\ref{sopt}) for the pole state and for the decaying state, while readjusting the real part of the latter so as to fit the 75 MeV binding. For the convenience of the ATMS three-body calculation,\cite{Akaishi86} the obtained optical interactions are simulated with Gaussian local potentials (units in MeV and fm) as 
\begin{eqnarray}
v_1^{\rm{opt}} = (-669 - i120) ~{\rm{exp}} (-(\frac{r}{0.66})^2), \\
v_1^{\rm{opt}} = (-659 - i~ 60) ~{\rm{exp}} (-(\frac{r}{0.66})^2)~~ 
\label{gauss}
\end{eqnarray}
for the pole state and for the intrinsic decaying state, respectively. It should be noted that the Gaussian range parameter we use (0.66 fm) corresponds to a Yukawa range parameter of 0.66/2 = 0.33 fm, which is not so different from the value adopted in Ref.\cite{Shevchenko07a}. The obtained width of $K^-pp$ is 100 MeV (the setting value) for the pole state and 54 MeV for the intrinsic decaying state. This gives a reasonable account for the difference between Shevchenko {\it et al}.'s 100 MeV width and Y-A's 60 MeV width.

\section{Conclusions}

In Section 2 we treated the two-body $K^-p$ system starting from the coupled channels of $K^-p$ and $\pi\Sigma$ and clarified that the pole-state solution leads to unphysical behaviors when the pole state approaches the $\pi\Sigma$ emission threshold. This difficulty can be avoided by taking into account the fact that the emitted $\pi$ and $\Sigma$ are on-shell particles with real energies. We have shown that the experimentally observed resonance is not the "pole state", but the "decaying state" introduced by Kapur and Peierls,\cite{Kapur38} and proposed the "intrinsic decaying state" as a representative of the exact $K^- p$ spectral shape. In Section 3 we discussed the problem of the $K^-pp$ width disagreement stated in Introduction. Shevchenko {\it et al}.'s width\cite{Shevchenko07b} is that of the pole state of $K^-pp$ and is not directly related to the experimental observation, while Y-A's one\cite{Yamazaki07b} is close to the intrinsic decaying state, which corresponds to the experimental shape of the missing-mass spectrum. It is concluded that {\it the treatment with the energy-independent complex $\bar KN$ interaction is a suitable means to obtain the intrinsic decaying state of $K^-pp$}. Shevchenko {\it et al}.'s statement\cite{Shevchenko07a} that "because the coupling of the two-body $K^-p$ channel to the absorptive $\pi Y$ channels was substituted by an energy-independent complex $\bar KN$ potential, Y-A's results for the binding energy and width of the $K^-pp$ system provide at best only a rough estimate", is a superficial view. 

\section*{Acknowledgments}

The authors thank Professors P. Kienle, M. Kawai, O. Morimatsu and K. Yazaki for stimulating and valuable discussions. They acknowledge the receipt of Grant-in-Aid for Scientific Research of Monbu-Kagakusho of Japan.

\section*{Appendix}

The coupled-channel equation for the radial wave functions, $u_1(r)$ and $u_2(r)$, of the present interaction model is written as follows: 
\begin{eqnarray}
- \frac {d^2}{dr^2} u_1(r) + (G_{11}+G_{12}) e^{-\Lambda r} = \kappa_1^2~u_1(r), 
\label{ccr1} \\
- \frac {d^2}{dr^2} u_2(r) + (G_{21}+G_{22}) e^{-\Lambda r} = \kappa_2^2~u_2(r),
\label{ccr2}
\end{eqnarray} 
where 
\begin{equation}
G_{ij} = 2 s_{ij} \sqrt{\frac {\mu_i}{\mu_j}} \Lambda^3 \int_0^\infty dr' e^{-\Lambda r'} u_j(r'). 
\label{coef}
\end{equation} 
From Eqs.~(\ref{ccr1}) and (\ref{ccr2}) radial solutions with outgoing wave boundary conditions are obtained to be
\begin{eqnarray}
u_1(r)= \frac {G_{11}+G_{12}}{\Lambda^2+\kappa_1^2}\{e^{-\Lambda r}-e^{i\kappa_1 r}\}, 
\label{sol1} \\
u_2(r)= \frac {G_{21}+G_{22}}{\Lambda^2+\kappa_2^2}\{e^{-\Lambda r}-e^{i\kappa_2 r}\}. 
\label{sol2} 
\end{eqnarray} 
The consistency condition of Eq.~(\ref{coef}) between $u_i$'s and $G_{ij}$'s gives an eigen-value equation of 
\begin{equation}
\{ (\Lambda-i\kappa_1)^2+s_{11} \Lambda^2 \} \{ (\Lambda-i\kappa_2)^2+s_{22} \Lambda^2 \} = s_{12} s_{21} \Lambda^4, 
\label{eig}
\end{equation} 
which determines the pole energy. If an additional condition of on-shellness is imposed on $\kappa_2$, the intrinsic decaying state of Eq.~(\ref{decayfn1}) is obtained.

The radial wave function, $u_1(r)$, is also obtained from a single-channel equation with an optical potential, $s_1^{\rm{opt}}$:
\begin{equation}
- \frac {d^2}{dr^2} u_1(r) + G_1^{\rm{opt}} e^{-\Lambda r} = \kappa_1^2~u_1(r), 
\label{opt1}
\end{equation} 
where 
\begin{equation}
G_1^{\rm{opt}} = 2 s_1^{\rm{opt}} \Lambda^3 \int_0^\infty dr' e^{-\Lambda r'} u_1(r'). 
\label{copt}
\end{equation} 
The solution with an outgoing wave boundary condition is obtained to be
\begin{equation}
u_1(r)= \frac {G_1^{\rm{opt}}}{\Lambda^2+\kappa_1^2} \{ e^{-\Lambda r}-e^{i\kappa_1 r} \}, 
\label{solopt} \\
\end{equation} 
The consistency condition of Eq.~(\ref{copt}) between $u_1$ and $G_1^{\rm{opt}}$ gives an eigen-value equation of 
\begin{equation}
 (\Lambda-i\kappa_1)^2+s_1^{\rm{opt}} \Lambda^2 = 0. 
\label{eigopt}
\end{equation} 
This equation gives the eigen-value of Eq.~(\ref{E0}),
\begin{equation}
E = \frac {\hbar^2}{2\mu_1} \kappa_1^2 = -\frac {\hbar^2}{2\mu_1} \Lambda^2 \{ \sqrt{-s_1^{\rm{opt}}} - 1 \}^2. 
\label{Eopt}
\end{equation} 
Since $s_1^{\rm{opt}}$ is a complex number, the phase of $\sqrt{-s_1^{\rm{opt}}}$ is uniquely determined when a proper Riemann's sheet is assigned. A pole on the $\bar KN~ [+]$ sheet of positive Im $\kappa_1$ is the quasi-bound state pole, and another pole on the $\bar KN~[-]$ sheet of negative Im $\kappa_1$ is the resonance or virtual-state pole, roughly speaking, depending on the sign of Re $E$. In the case of Eq.~(\ref{strength}), for example, the energy is obtained to be $E = -27 - i\,20$ MeV ($\bar KN$ quasi-bound state) with $s_1^{\rm{opt}} = -1.393 - i\, 0.142$ on the $\bar KN$ $[+]$ observable sheet, whereas it is  $E = -3946 - i \,0$ MeV ($\bar KN$ virtual state) with $s_1^{\rm{opt}} = -1.156 - i \,0$ on the $\bar KN$ $[-]$ sheet. Note that the former "$\Lambda(1405)$" pole is the quasi-bound state pole with respect to $\kappa_1$ and, at the same time, is the Feshbach resonance pole with respect to $\kappa_2$ as discussed in Fig.~\ref{Poles} and the text. The latter pole lies very far from the observation axis.

Rewriting Eq.~(\ref{eig}) as 
\begin{equation}
(\Lambda-i\kappa_1)^2+s_{11} \Lambda^2 - s_{12} \frac {\Lambda^4} {(\Lambda-i\kappa_2)^2+s_{22} \Lambda^2} s_{21} = 0,  
\end{equation} 
and comparing it with Eq.~(\ref{eigopt}), we obtain the relation of Eq.~(\ref{sopt}), 
\begin{equation}
s_{1}^{\rm {opt}} = s_{11}-s_{12} \frac {\Lambda^2}{(\Lambda-i\kappa_2)^2+s_{22}\Lambda^2} s_{21} 
\label{eig2}
\end{equation}
without any approximation.
Eq.~(\ref{eig2}) means that a loop integral of Green's function in Channel 2 becomes 
\begin{equation}
\int \int d\vec q' d\vec q ~g(\vec q') \langle \vec q' \mid \frac{1}{E-h_{22}+i\epsilon} \mid \vec q \rangle ~g(\vec q) = -(\pi^2 \frac {2 \mu_2}{\hbar^2} \Lambda) \frac {\Lambda^2} {(\Lambda-i\kappa_2)^2+s_{22} \Lambda^2}. 
\label{loop2}
\end{equation} 
Similarly, a loop integral of Green's function in Channel 1 is evaluated as
\begin{equation}
\int \int d\vec q' d\vec q ~g(\vec q') \langle \vec q' \mid \frac{1}{E-h_{1}^{\rm{opt}}(E)+i\epsilon} \mid \vec q \rangle ~g(\vec q) = -(\pi^2 \frac {2 \mu_1}{\hbar^2} \Lambda) \frac {\Lambda^2} {(\Lambda-i\kappa_1)^2+s_1^{\rm{opt}} \Lambda^2}. 
\label{loop1}
\end{equation} 
This formula is used to derive a Breit-Wigner type spectrum from Eq.~(\ref{cross}), where the $t$-matrix is separable, $\langle \vec k \mid t_{ij} \mid \vec q \rangle = g(\vec k)~ T_{ij} ~g(\vec q)$, in the present model.

\end{document}